\documentclass[aps,showpacs,preprintnumbers,amsmath,amssymb,twocolumn]{revtex4}
\UseRawInputEncoding
\usepackage{dcolumn}
\usepackage[dvips]{epsfig}
\usepackage{float}
\usepackage{graphics,epsfig}
\usepackage{graphicx}
\usepackage{dcolumn}
\usepackage{bm}
\usepackage{gensymb}
\usepackage{footnote}
\usepackage{booktabs}
\usepackage{subfigure}
\usepackage{epstopdf}
\usepackage{aasmacros}
\usepackage{amsmath,amssymb,amsfonts}

\begin{document}
\baselineskip=0.4cm
\title{\bf  Black hole images: A Review}

\author{Songbai Chen $^{1,2}$ \footnote{Corresponding author:  csb3752@hunnu.edu.cn}, Jiliang Jing $^{1,2}$, Wei-Liang Qian $^{2,3,4}$,  Bin Wang $^{2,5}$ }
\affiliation{$ ^1$ Department of Physics, Synergetic Innovation Center for Quantum Effects and Applications,
Hunan Normal University, Changsha, Hunan 410081, People's Republic of China\\
$ ^{2}$
Center for Gravitation and Cosmology, College of Physical Science and Technology, Yangzhou University, Yangzhou 225009, People's Republic of China\\
$ ^{3}$  Escola de Engenharia de Lorena, Universidade de S\~{a}o Paulo, 12602-810, Lorena, SP, Brazil\\
$ ^{4}$ Faculdade de Engenharia de Guaratinguet\'{a}, Universidade Estadual Paulista, 12516-410, Guaratinguet\'{a}, SP, Brazil\\
$ ^{5}$ School of Aeronautics and Astronautics, Shanghai Jiao Tong University, Shanghai 200240, People's Republic of China}

\begin{abstract}
\baselineskip=0.35 cm

In recent years, unprecedented progress has been achieved regarding
black holes' observation through the electromagnetic channel.
The images of the supermassive black holes M87$^{*}$ and Sgr A$^{*}$ released by
the Event Horizon Telescope (EHT) Collaboration provided direct visual
evidence for their existence, which has stimulated further studies on
various aspects of the compact celestial objects.
Moreover, the information stored in these images provides a new way to
understand the pertinent physical processes that occurred near the
black holes, to test alternative theories of gravity, and to furnish insight
into fundamental physics. In this review, we briefly summarize the recent developments on the topic.
In particular, we elaborate on the features and formation mechanism of
black hole shadows, the properties of black hole images illuminated by
the surrounding thin accretion disk, and the corresponding
polarization patterns. The potential applications of the relevant studies are also addressed.

\end{abstract}

 \maketitle
\newpage
\section{Introduction}

The releasing of the first image of the supermassive black hole M87$^{*}$  by the EHT Collaboration in 2019 \cite{2019ApJ...875L...1E,2019ApJ...875L...2E,2019ApJ...875L...3E,2019ApJ...875L...4E,2019ApJ...875L...5E,2019ApJ...875L...6E} is a milestone event in physics. It provided direct visual evidence of black hole in our universe, which means that black hole is no longer just a theoretical model. Combining with the recently published black hole image of Sgr A$^{*}$ \cite{2022APJ...930L...17E}, it is widely believed that the observational  astronomy of black holes has entered a new era of rapid progress.

One of the most important ingredients  in the images is the black hole shadow \cite{Falcke:1999pj}. It is a two dimensional dark region in the observer's sky, which is caused by light rays falling into
an event horizon of a black hole \cite{10.1093...mnras...131.3.463,Bardeen:1973tla,Cunningham:1975zz,Chandrasekhar:1984siy}. The captured light rays are very close to the black hole so that the shape and size of
the shadow carry the fingerprint of the celestial object. Therefore,  the study of shadows is beneficial to identify black holes, to examine theories of gravity including general relativity, and to further understand some fundamental problems in physics.

From the light propagation in spacetime,  black hole shadow  depends on the light source, the background spacetime and even the properties of electrodynamics obeyed by photon itself.  The light sources in many theoretical investigations are assumed to homogeneously distribute  in the total celestial sphere. However, in the real astronomical environment, accretion disk is a kind of actual and feasible light
sources around black holes due to its electromagnetic emission. Undoubtedly, the matter configuration and the accretion process in the disks are indelibly imprinted on black hole images. Conversely, analyzing
the luminosity distribution and electromagnetic signals stored in the images can extract the information about the matter fields and the physical processes near the black holes. The twisting patterns in the first polarized image of the M87* black hole \cite{2021ApJ...910L..12E,2021ApJ...910L..13E} revealed the presence of a poloidal magnetic field about $1\sim 30G$ near the black hole. Thus,
black hole images with  polarized information  provide another new way to probe  the matter distribution,  the electromagnetic interactions and the accretion processes in the strong gravity region of black holes.

The literature related to black hole images is rapidly increasing. Therefore, it is necessary to summarize the existing works at the present time and make prospects for the future.
This review focuses on the black hole images, and their rapid development and potential applications. We first introduce the basic concepts of black hole shadows and summarize the main features of the shadows and their formation mechanism, and review how the black hole shadows are determined by the fundamental photon orbits \cite{Cunha:2017eoe} and the corresponding invariant manifolds \cite{Grover:2017mhm}. Next, we introduce the images of the black holes surrounded by a thin disk and their polarization patterns arising from synchrotron radiation. We also discuss the aspects of the potential applications of black hole images.

\section{Black hole shadows and their formation}
\begin{figure}
\includegraphics[width=5cm ]{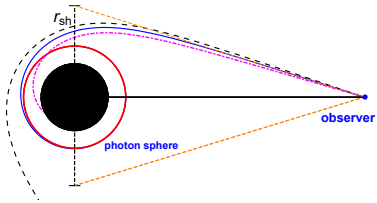}
\caption{The event horizon (the black disk), the photon sphere $r_{\rm ps}=3M$ (the red circle) and the shadow with a radius $r_{\rm sh}=3\sqrt{3}M$ for a Schwarzschild black hole \cite{Wang:2022kvg}.}
\label{pst}
\end{figure}

It is well known that the shadow of a common object is determined by the light rays passing through the edge of the object. However, for the objects with strong gravity, such as the black hole, the situation is different.
According to general relativity, light rays travelling in a black hole spacetime can be deflected due to the gravitational field of the black hole. This phenomenon is known as gravitational lensing, which is analogous to optical
lensing \cite{Perlick:2004tq, Blandford:1991xc,Refsdal:1993kf}. The deflection angle of the light ray increases with the decreasing of its impact parameter. Therefore, it is easy to infer that the light rays passing very close to  the black hole will be captured. Black hole shadow is a dark silhouette observed in the sky  originating from these captured light rays.
Although the black hole shadow is caused by the photons fallen into the event horizon, its size is not equal to that of this null hypersurface. Actually, for a Schwarzschild black hole, its shadow is about 2.5 times as large as the event horizon in angular size \cite{10.1093...mnras...131.3.463,Bardeen:1973tla}. This can be explained by two reasons. Firstly, it is not only the photons near the event horizon can be captured by a black hole. In fact, there exists a photon sphere outside the event horizon, which is an envelope surface of unstable photon circular orbits in the spacetime \cite{Cunha:2018acu,Perlick:2021aok,Wang:2022kvg}. The light rays  entered the photon sphere will be captured by the black hole as shown in Fig.\ref{pst}. Thus, the boundary of the shadow is determined by the photon sphere rather than the event horizon. Secondly, due to the strong bending of light rays induced by black hole's gravity, both the size and shape of the observed dark shadow are different from those naively based on Euclidean geometry without gravity.

\subsection{Features of black hole shadows}
The shape and size of black hole shadows depend on the black hole parameters and the observer inclination. For a Schwarzschild black hole, the shadow is a perfect disk for the observer with arbitrary inclination \cite{Bardeen:1973tla,Chandrasekhar:1984siy}. For a rotating Kerr black hole, the shadow also presents a
circular silhouette for the observer located on its rotation axis. However, for the observer in the equatorial plane,  the shadow gradually becomes a ``D"-shaped silhouette with increasing black hole spin \cite{Bardeen:1973tla,Chandrasekhar:1984siy}. For a Konoplya-Zhidenko rotating non-Kerr black hole with an extra deformation parameter described the deviations from the Kerr metric, the ``D"-shaped shadow could disappear and the special cuspy shadow could emerge in a certain range of parameter values for the equatorial observer \cite{Wang:2017hjl}. This is also
true for black holes with Proca hair \cite{Cunha:2017eoe,Sengo:2022jif}.
Thus, the dependence of shadows on black hole parameters could provide a potential tool to identify black holes in nature. It also triggers the further study of black hole shadows in various theories of gravity \cite{Amarilla:2010zq,Dastan:2016vhb,Long:2020wqj,Stepanian:2021vvk,Prokopov:2021lat,Younsi:2021dxe}.

Black hole shadow also depends on the (non-)integrability of motion equation of photon travelling in the background spacetime. The completely integrable systems are such kind of dynamical systems where the number of the first integrals is equal to its degrees of freedom.
Generally, in static spacetimes of black holes with spherical symmetry, such as in the Schwarzschild black hole spacetime, the dynamical system of photons is completely integrable since it possesses three independent first integrals, i.e., the energy $E$, the $z$-component of the angular momentum $L_{z}$ and the Carter constant $Q$ \cite{Carter:1968rr}. This ensures the motion of photons is regular so that black hole shadow has the same shape as the photon sphere
\begin{figure}
\center{\includegraphics[width=4cm ]{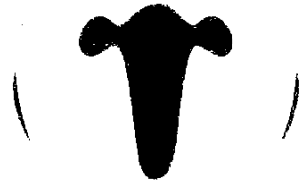}
\caption{The eyebrowlike shadows near the primary shadow for a Kerr black hole with scalar hair \cite{Cunha:2015yba}.}
\label{khair}}
\end{figure}
surface. In these completely integrable systems, the black hole shadow can be calculated by analytical methods \cite{Bardeen:1973tla,Chandrasekhar:1984siy}.
However, as the dynamical system of photons is not completely integrable, the null geodesic equations are not be variable
separable because there is no existence of a Carter-like constant apart from the usual two integrals of motion $E$ and $L_z$. This
implies that the motion of photons could be chaotic and sharply affect the shadow so that its shape is different from that of the photon sphere surface \cite{Cunha:2015yba,Vincent:2016sjq,Cunha:2016bjh,Wang:2017qhh,Wang:2018eui,Wang:2019tjc}. In particular, due to chaotic lensing, there are eyebrowlike shadows  with the self-similar fractal structure near the primary shadow, as shown in Fig.\ref{khair}. In addition, the black hole shadows in the nonintegrable cases can be only obtained by numerical simulations with the so-called ``ray-tracing" codes \cite{Bohn:2014xxa,Cunha:2015yba,Wang:2017qhh,Riazuelo:2015shp}.

\subsection{Formation mechanism of black hole shadows}

The photon sphere plays an important role in the formation of black hole shadows. Actually, the  photon sphere is composed of unstable photon circular orbits around black holes. The unstable photon circular orbits in the equatorial plane are determined by the effective potential and its derivatives \cite{Claudel:2000yi,Virbhadra:2002ju}, i.e., $V_{\rm eff}(r)=0$, $V_{{\rm eff}} (r)_{,r}=0$ and $V_{{\rm eff}}(r)_{,rr}<0$. These unstable orbits
can also be obtained by a geometric way with Gauss curvature and geodesic curvature in the optical
geometry \cite{Qiao:2022ifk}. For a static four dimensional spacetime, its optical geometry  restricted in the equatorial plane is defined by ${\rm d}t^2\equiv g^{\rm OP}_{ij}{\rm d}x^i{\rm d}x^j$, $i=r, \phi$. The geodesic curvature and Gauss curvature can be expressed as \cite{Qiao:2022ifk}
\begin{eqnarray}
&&\kappa_{\rm geo}=\frac{1}{2\sqrt{g^{\rm OP}_{rr}}}\frac{\partial \ln g^{\rm OP}_{\phi\phi}}{\partial r},\\
&&\kappa_{\rm Gau}=-\frac{1}{\sqrt{g^{\rm OP}}}\bigg[\frac{\partial}{\partial \phi}\bigg(\frac{1}{\sqrt{g^{\rm OP}_{\phi\phi}}}\frac{\partial \sqrt{ g^{\rm OP}_{rr}}}{\partial \phi}\bigg)\nonumber\\
&&+\frac{\partial}{\partial r}\bigg(\frac{1}{\sqrt{g^{\rm OP}_{rr}}}\frac{\partial \sqrt{ g^{\rm OP}_{\phi\phi}}}{\partial r}\bigg)\bigg].
\end{eqnarray}
The geodesic curvature $\kappa_{\rm geo}=0$ gives the radius of the circular photon orbit and the positive (negative) of Gauss curvature $\kappa_{\rm Gau}$ determines that the circular photon orbit is stable (unstable).

\textit{ Fundamental photon orbits}\;\; The unstable photon circular orbits in the equatorial plane are often called light rings. Light rings also affect dynamical properties of ultracompact objects \cite{Cunha:2017eoe}. For the ultracompact objects without horizon \cite{Cunha:2017qtt}, light rings often come in pairs, one stable and the other unstable. The existence of a stable light ring always implies a spacetime instability \cite{Cunha:2022gde}.
In the Schwarzschild spacetime, light rings are the only bound photon orbits.  In the rotating Kerr spacetime, there are two light rings located in the equatorial plane, one for co-rotating photon and one for counter-rotating photon with respect to the black hole. Moreover, there  also exist the non-planar bound photon orbits with constant
$r$ and motion in $\theta$, known as spherical orbits ( see also Fig. 2 in \cite{Cunha:2017eoe}). These spherical orbits are unstable and completely determine the Kerr black hole shadow.

Fundamental photon orbits are the generalization of light rings and spherical orbits
in usual stationary and axisymmetric spacetimes \cite{Cunha:2017eoe}. The definition of fundamental photon orbits is given in \cite{Cunha:2017eoe}.
According to the features of orbits, the fundamental photon orbits can be categorized as $X^{n_r \pm}_{n_{s}}$, where $X=\{O,C\}$, and $n_r,n_{s}\in \mathbb{N}_0$.
\begin{figure}
\includegraphics[width=6cm]{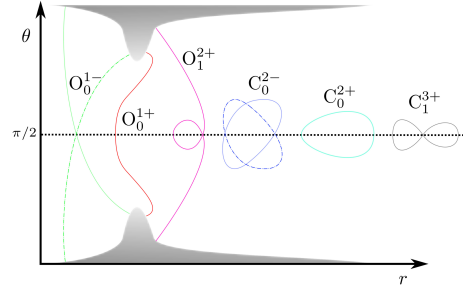}
\caption{Some fundamental photon orbits in the $(r, \theta)$-plane and their classification. The grey areas represent forbidden regions for the effective potential \cite{Cunha:2017eoe}.}
\label{fporbit}
\end{figure}
The orbit $O$ is open and it can reach the boundary of the effective potential. The orbit $C$ is closed and it can not reach the boundary. The sign $+(-)$ denotes the even (odd) parity of the orbit under the  $\mathbb{Z}_2$ reflection symmetry around
the equatorial plane. $n_r$ is the number
of distinct $r$ values at where the orbit crosses the equatorial plane. For the light rings lied in the equatorial plane, their $n_r=0$ because such special kind of orbits never cross the equatorial plane. $n_{s}$ is the number of self-intersection points of the orbit. In Fig.\ref{fporbit}, some fundamental photon orbits and their classification are illustrated in the $(r, \theta)$-plane.

\begin{figure}
 \subfigure[]{ \includegraphics[width=6cm ]{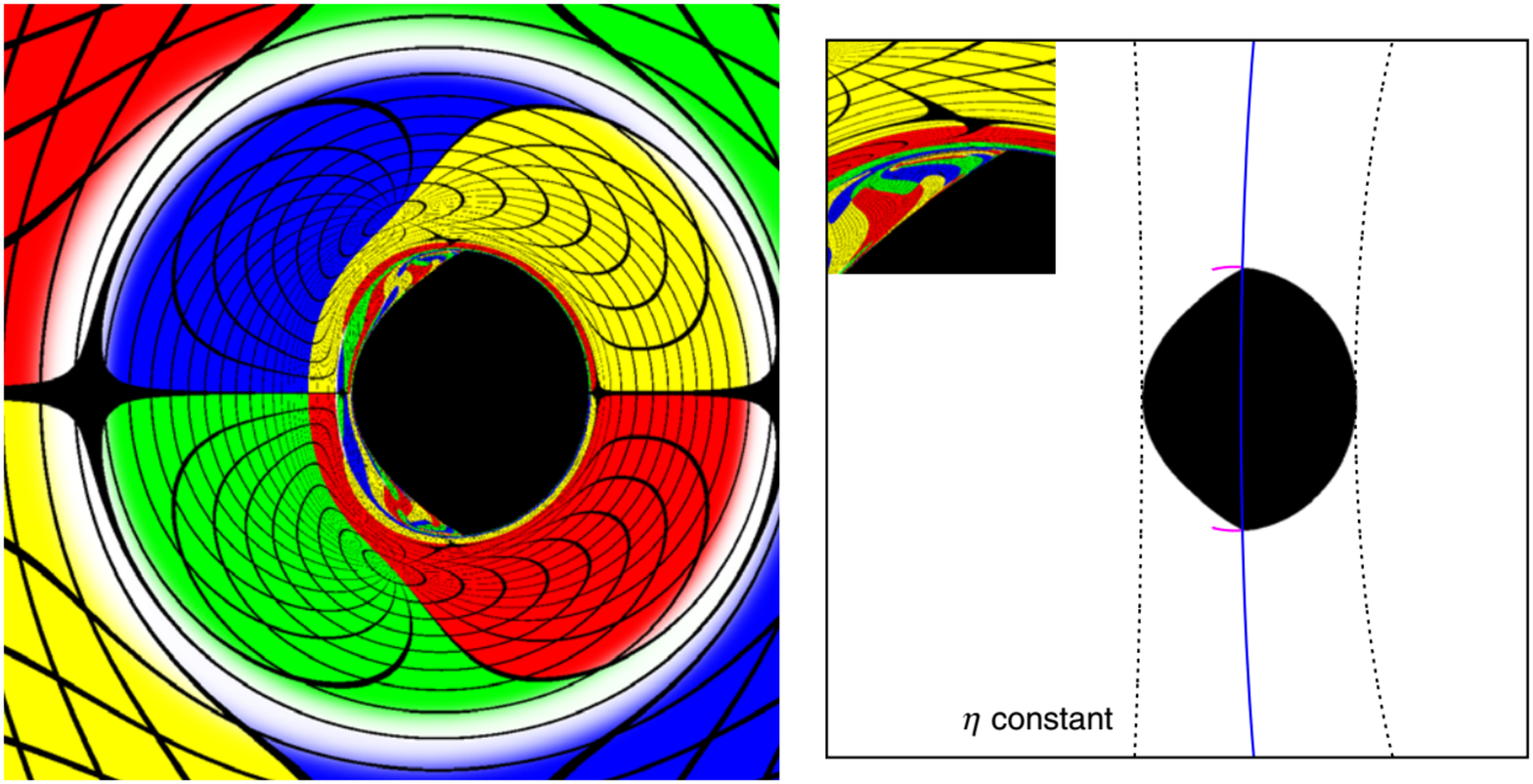}}\\
 \subfigure[]{ \includegraphics[width=7cm ]{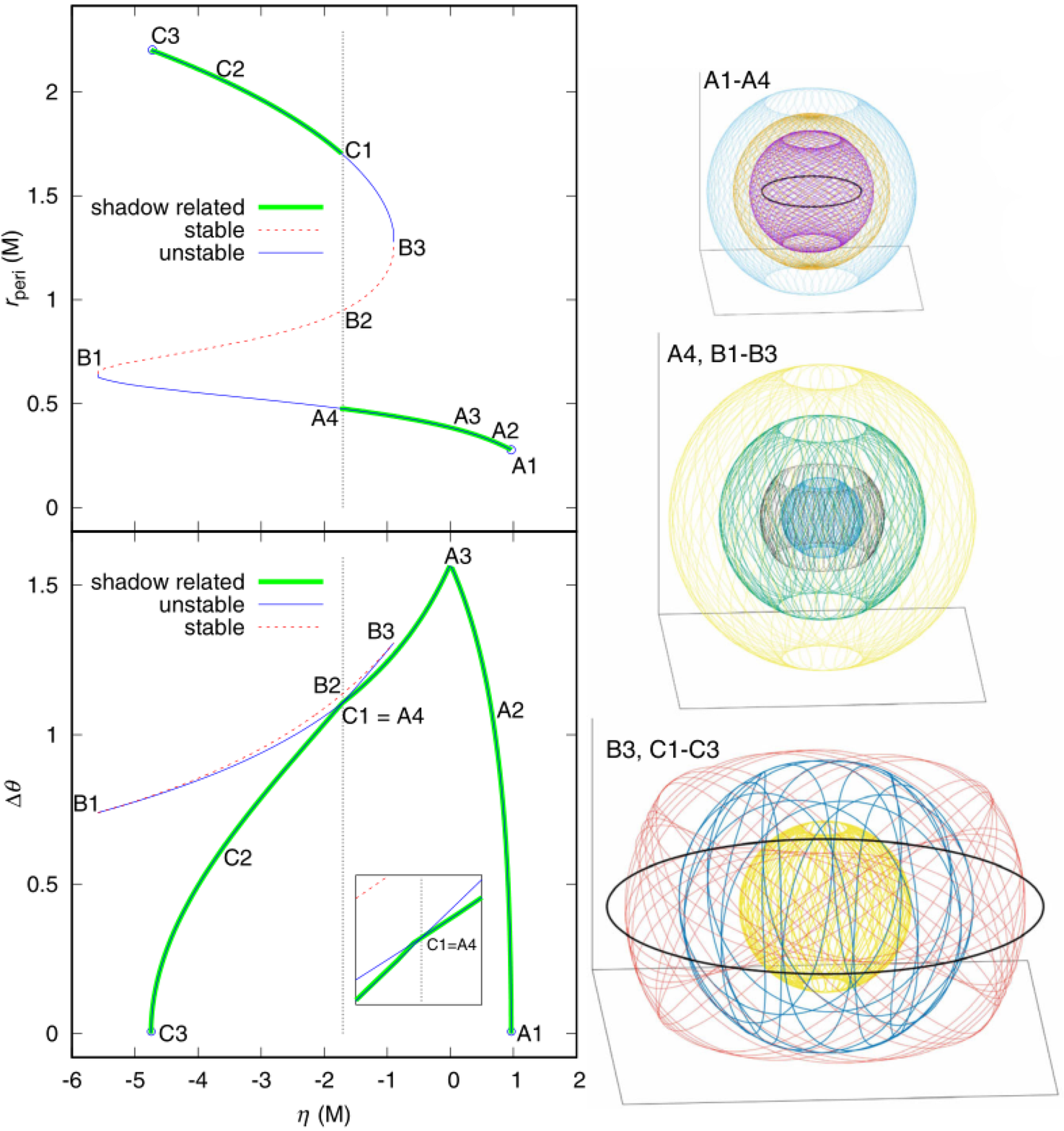}}
\caption{ The cuspy shadow (a) and the fundamental photon orbits (b) for the Kerr black hole with Proca hair\cite{Cunha:2017eoe}.}
\label{ssab}
\end{figure}

By making use of the fundamental photon orbits, P. V. P. Cunha et al. \cite{Cunha:2017eoe} explained the formation of the cuspy silhouette of a Kerr black hole with Proca hair shown in Fig.\ref{ssab}(a). To clearly demonstrate  the formation mechanism of the black hole shadow, ten fundamental photon orbits are selected out and marked by ``A1-A4, B1-B3, C1-C3". Then, the distribution of $\Delta\theta\equiv|\theta_{\text{max}}-\frac{\pi}{2}|$ and $r_{\text{peri}}$ with the impact parameter $\eta$ are presented for each fundamental photon orbit, where $\theta_{\text{max}}$ is the maximal/minimal angular coordinate at the spherical orbit and $r_{\text{peri}}$ is the perimetral radius as a spherical orbit crosses the equatorial plane. The right panel of Fig.\ref{ssab}(b) shows the spatial trajectories of the ten fundamental photon orbits in Cartesian coordinates, which move around the black hole. The orbits A1 and C3 are the unstable prograde and retrograde light rings respectively shown as two black circles on the equatorial plane. Other
fundamental photon orbits are non-planar bound photon orbits crossing the equatorial plane. The continuum of fundamental photon orbits can be split into one stable branch (the red dotted line) and two unstable branches (the green and blue lines). The swallow-tail shape pattern related to the fundamental photon orbits in the $\eta-\Delta\theta$ plane yields a jump occurred at A4 and C1 in the $\eta-r_{\text{peri}}$ plane. The discontinuity  originating from this jump, i.e., $r_{\text{peri}(\text{C1})}>r_{\text{peri}(\text{A4})}$, induces the emergence of the cuspy shadow.  The unstable
fundamental photon orbits (C1-B3) non-related
to the shadow could be associated to a set of lensing
patterns attached to the shadow edge, called ``eyelashes"
in Fig.\ref{ssab}(a). In the black hole spacetimes where
there exists a second pair of
light rings, the fundamental photon orbits can be classified into two fully disconnected branches: the shadow related branch
and the non-shadow related one. If the shadow related branch is connected, the shadow edge will be smooth with no cusp. But the eyelashes, caused by the non-shadow related unstable branch, appear to be disconnected from the shadow, forming a pixelated banana-shaped strip in the lensing image as shown in  Fig.\ref{khair}. This feature has been dubbed ``ghost shadow" in \cite{Sengo:2022jif}. This mechanism is also applied to explain the formation of the cuspy shadow in the Konoplya-Zhidenko rotating non-Kerr black hole spacetime \cite{Wang:2017hjl}.
It is further confirmed that the unstable fundamental photon orbits play an important role in determining the boundary of shadow and the patterns of the shadow shape. In terms of a toy model \cite{Qian:2021qow}, the feature of cuspy shadow can be derived by
employing the Maxwell construction for phase transition in a
two-component system. In addition, the shadows for the black holes with general parameterized metrics
 have been studied by using the fundamental photon orbits  \cite{Li:2021mnx,Bambi:2015rda,Ghasemi-Nodehi:2015raa,Kumar:2018ple,Medeiros:2019cde,Carson:2020dez}. These studies could also be beneficial to test the Kerr hypothesis through black hole shadows.

\textit{Invariant phase space structures} The invariant phase space structures are very important for dynamical systems because they remain invariant under the dynamics. There are several types of invariant structures including fixed points, periodic orbits and invariant manifolds. The simplest is the fixed points. These phase space structures are applied extensively to design space trajectory for various of spacecrafts \cite{2009CeMDA.105...61M,2001CeMDA..81...63K,1997gfrt.book.....H,1980CeMec..22..241R}. Since black hole shadow depends on the dynamics of photons in the background spacetime, the invariant phase space structures should also play an important role in the formation mechanism of the shadow. For a dynamical system of photons travelling in a curved spacetime, its fixed points can be determined by the conditions
\begin{eqnarray}
\label{bdd}
\dot{x}^{\mu}=\frac{\partial H}{\partial p_{\mu}}=0,\;\;\;\;\;\;\;\;\;\;\;\;\;\;
\dot{p}_{\mu}=-\frac{\partial H}{\partial x^{\mu}}=0,
\end{eqnarray}
where $q^{\mu}=(t,r,\theta,\varphi)$ and $p_{\nu}=(p_{t},p_{r},p_{\theta},p_{\varphi})$, $H$ is the Hamiltonian of the system. Actually, the light rings  on the equatorial plane  are the fixed points for the photon motion \cite{Grover:2017mhm,Cunha:2017eoe}. In the vicinity of the fixed points, one can linearize the equations (\ref{bdd}) and obtain a matrix equation
\begin{eqnarray}
\label{xxh}
\mathbf{\dot{X}}=J\mathbf{X},
\end{eqnarray}
where $\mathbf{X}=(q^{\mu},p_{\nu})$ and $J$ is the Jacobian matrix. The eigenvalues $\mu_{j}$ of the Jacobian matrix $J$ determine the local dynamical properties of the system
near the fixed points (see also in Fig.\ref{lx0}). The stable and unstable invariant manifolds correspond to the cases of ${\rm Re}(\mu_{j})<0$ and ${\rm Re}(\mu_{j})>0$, respectively; while the center manifold corresponds to the case with ${\rm Re}(\mu_{j})=0$ where the eigenvalues $\mu_{j}$ are pure imaginary numbers. Due to the special properties of the invariant manifolds, there is no trajectory crossing the invariant manifolds. Points in the unstable (stable) invariant manifold  move to the fixed points exponentially in backward (forward) time.
In terms of Lyapunov's central limit theorem, the eigenvalue with ${\rm Re}(\mu_{j})=0$ leads to the so-called Lyapunov orbits, which is
a one-parameter family $\gamma_{\epsilon}$ of periodic orbits \cite{2009CeMDA.105...61M,2001CeMDA..81...63K,1997gfrt.book.....H,1980CeMec..22..241R,Grover:2017mhm}. These orbits $\gamma_{\epsilon}$ in the center manifold collapse into a fixed point as $\epsilon\rightarrow0$. Similarly, the periodic orbits also have their own stable and unstable manifolds. These invariant phase space structures are also shown in Fig. 1 in \cite{Grover:2017mhm}. Obviously, the photon spheres and other periodic orbits can be generalized to the Lyapunov orbits related to the fixed points \cite{Grover:2017mhm}.

\begin{figure}
\includegraphics[width=5cm ]{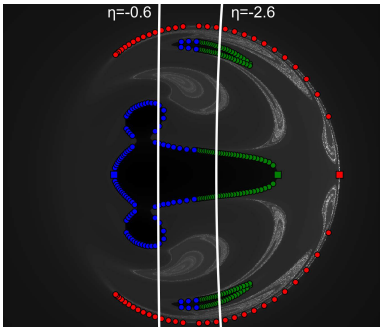}
\caption{Intersections of the unstable manifolds of $\mathfrak{L}_1$, $\mathfrak{L}_2$, and $\mathfrak{L}_3$ as well as their Lyapunov orbits with the image plane. Lyapunov orbits related to $\mathfrak{L}_1$, $\mathfrak{L}_2$, and $\mathfrak{L}_3$ are marked by red, green, and blue dots, respectively \cite{Grover:2017mhm}.}
\label{lx}
\end{figure}

These concepts in dynamical systems provide a powerful theoretical foundation for understanding the formation of shadows cast by black holes.
The unstable invariant manifold builds a bridge between the photon sphere and the observer because such manifold  can approach the fixed points exponentially in backward time \cite{Grover:2017mhm,Wang:2017qhh}.  Only the Lyapunov family of the spherical orbits near the unstable fixed points are responsible for generating the black hole shadow. For a Kerr black hole with scalar hair, there are three unstable light rings $\mathfrak{L}_1$, $\mathfrak{L}_2$ and $\mathfrak{L}_3$. However, the Lyapunov orbits associated with $\mathfrak{L}_1$ is non-spherical and only the orbits emanating from $\mathfrak{L}_2$ and $\mathfrak{L}_3$ are spherical \cite{Grover:2017mhm}. The numerical simulated shadow in Fig.\ref{lx}
shows that the complicated and disconnected boundaries of the shadow are completely determined by the Lyapunov spherical orbits. Thus,
the invariant manifolds of certain Lyapunov orbits are directly related to black hole shadows even in the case of complicated non-convex, disconnected shadows. Moreover, the curved streamlines in the unstable invariant manifolds could lead to that the shape of the black hole shadow detected by the observer at spatial infinity differs from that of the photon sphere surface \cite{Grover:2017mhm,Wang:2017qhh,Wang:2019tjc}.

\section{Image of a black hole with a thin accretion disk}

In the real astrophysical conditions, a black hole is surrounded by a hot accretion disk within which it emits a characteristic spectrum of electromagnetic radiation.  The electromagnetic radiation emitted by the disk illuminates the background around the black hole and makes the black hole shadow visible.  Thus, the accretion disk is the actual light source in the formation of  black hole shadows. Clearly, the black hole image cast by the light source with the disk-like structure differs from that by the former homogeneous light source in the previous analyses. Simultaneously, due to the strong gravitational lensing near the central black hole, the shape of the accretion disk is heavily distorted.

Luminet first simulated a photograph of a Schwarzschild black hole with a rotating thin accretion disk \cite{1979A&A....75..228L}. Here,
the proper luminosity of the disk is calculated according to the model described by Page and Thorne \cite{Page:1974he} in its relativistic version, where the intensity of radiation emitted at arbitrary given point of the disk only depends on the radial distance to the black hole.
As shown in Fig.11 in \cite{1979A&A....75..228L}, the flying-saucer-shaped bright region is the primary image of disk, which is formed by the light emitted directly from the upper side of the disk \cite{1979A&A....75..228L}. Due to the considerable distortion caused by the strong gravitational lensing near the central black hole, the primary image related to the back part of the disk is completely visible rather than hidden by the black hole.

Moreover, one also sees a highly deformed image associated with the bottom of the gaseous disk. This is because the light rays emitted from the bottom side can climb back to the top and reach to the observer at the spatial distance \cite{1979A&A....75..228L}. Actually, the gravitational lensing gives rise to an infinity of images of the disk, which are caused by the light rays traveling around the black hole any number of times before reaching a distant astronomer  \cite{1979A&A....75..228L,2022PhRvD.105f4040B}. The number of times of light ray crossing the disk determines the order of the image. The higher order images are closer to the central black spot and become thinner and fainter. The inner infinite order image is related to the photon sphere, which represents the actual shadow boundary of the  black hole. Generally, it is difficult to distinguish  the higher order images optically because they are standing quite closely to each other. The central black area is the black hole shadow formed by the gravitational lensing and capture of light rays.

The existence of a dark gap between the primary image and the higher order images is not surprising because the accretion disk is forbidden to touch the surface of the black hole and then there is not any radiation from the region between the black hole's event horizon and the inner edge of the disk. Moreover, below the inner stable circular orbit, the disk is unstable so that the gas particles plunge directly towards the black hole without having enough time to emit electromagnetic radiation \cite{1979A&A....75..228L}. Unlike the shadow itself, the darkness in these
patches is of a fundamentally different nature, which may be filled with
the emission from lensed images of distant sources in the entire universe although
it will also be extremely faint \cite{Bronzwaer:2021lzo,Bronzwaer:2020vix}.

For a disk around the black hole, the region closer to the horizon is generally brighter because the gas is hotter there.
However, the apparent luminosity of the disk's image for the distant observer is very different from the intrinsic luminosity in the disk.
The main reason is that the electromagnetic radiation detected at a great distance undergoes shifts in frequency and intensity with respect to the original radiation emitted directly by the disk \cite{1979A&A....75..228L,James:2015yla}. There are two kinds of shift effects. One of them is the so-called gravitational redshift caused by the gravity of the central black hole, which lowers the frequency and decreases the intensity of the electromagnetic radiation.
The other is the well-known Doppler effect originating from the displacement of the source with respect to the observer.
Doppler effect gives rise to amplification for the approaching source and attenuation for the retreating source.
Therefore, for a disk rotating counterclockwise around the black hole, the apparent luminosity of the disk in the left side is brighter than that of in the right side \cite{1979A&A....75..228L}. The strong gravity of the black hole can give a speed of gas rotation close to the speed of light in the internal regions of an accretion disk, which yields a very strong difference of Doppler shift effects on two sides of the black hole.
This strong asymmetry of apparent luminosity is the main signature of the black hole image with a thin accretion disk. In short, the effects from Doppler shift and gravitational redshift drastically modify the luminosity distribution for the observed disk image at large distance.

The black hole spin hardly affects the shape of the primary image. The principal effect of black hole spin is to change the radius of the marginally stable orbit and hence to modify the location of the inner edge of the accretion disk.
Unlike in the case of a Schwarzschild black hole,  a rapid rotation of Kerr black hole could lead to that  the inner edge of the direct image coincides with  the higher order images, so the dark gap between them may no longer exist \cite{1993A&A...272..355V,1993Ap&SS.205..155V}.
Due to the inner edge of the accretion disk being located far deeper in the gravitational potential, the range of accessible redshift in the disk for the rapidly rotating Kerr black hole is far broader than for the Schwarzschild case.
Thus, the higher order images round a rapidly spinning black hole carry less flux than
in the Schwarzschild case, which means that they are much more difficult to spatially
resolve from the direct image of the disk in the rapidly rotating black hole case.

Moreover, the gravitational field of the accretion disk also affects the propagation of photon and further modifies the
shape of black hole shadow. Recently, a static axially symmetric solution,  which describes the superposition of a Schwarzschild black hole with a relativistic thin and heavy accretion disk ( Lemos-Letelier disk \cite{Lemos:1993qp}),  is applied to study black hole shadow \cite{2020JCAP...03..035C}.
This static disk with an inner edge is assumed to be made of two streams of counter-rotating
particles \cite{Lemos:1993qp}, which leads to a total vanishing angular momentum and ensures the existence of a static disk
in equilibrium with the black hole. A heavy accretion disk yields some new features for the black hole image \cite{2020JCAP...03..035C}.
There is a progressive optical enlargement of the disk image covering part of the shadow, despite the fact that the disk is infinitesimally thin. This is a consequence of the increasing light rays' bending towards the disk due to the increase of disk's ``weight". The heavy disk also stretches the black hole shadow so that there is an extra deformation of the shadow shape, which becomes more prolate as the disk contributes to a
higher fraction of the total mass. Furthermore, the nonintegrability of the photon motion arising from a heavy accretion disk also leads to some chaotic patterns both in the black hole shadow and the disk image. These features also appear in the gravity system of a Schwarzschild
black hole surrounded by a massive Bach-Weyl ring \cite{Wang:2019tjc}. The chaotic lensing also leads to some distinct differences in the shape of
photon sphere and the black hole shadow. This is because the chaotic orbits sharply modify the locally measured four-momentum of the photons reaching a distant observer and further influence the celestial coordinates of the images associated with these photons in the observer's sky, and the latter directly determines the shape of the black hole shadow and the disk image.

\section{Polarized image of a black hole}

Electromagnetic wave is a kind of transverse waves so the optical image of a black hole must carry the polarization information about the light emitted from the accretion disk around the black hole.
Recently, the EHT Collaboration has published the polarimetric image of the black hole M87$^{*}$  \cite{2021ApJ...910L..12E,2021ApJ...910L..13E}.
The twisting polarization patterns revealed the existence of magnetic field near the black hole. It is the first time to measure the polarization information characterized by the magnetic field near the black hole, which is helpful to understand the formation of the black hole jet far from 55 million light years.

Actually, in order to extract the information carried in the polarized image of a black hole, one must compare the observed polarimetry data with the theoretical one. Thus, it is very vital to make theoretical analyses and numerical simulations on the polarized images for various black holes.
In general, the polarization structures in the black hole images depend on
the details of the emitting plasma, principally the magnetic field geometry, and are also affected by the strongly curved
spacetime near the black hole. For the origin of the polarized emission
 around a black hole,  there is a typical scenario where the light with high polarization degrees, especially the linearly polarized light,
is produced by synchrotron emission in a compact and energetic
region of the inner hot disk \cite{Tsunetoe:2020pyz,Tsunetoe:2020nws}. It is because the relativistic Doppler beaming effect yields that the propagation directions of the photons emitted by a charged relativistic particle are beamed almost along the tangent direction of the particle's motion so that the light rays in the particle's orbital plane are linearly polarized. In the cold disk model \cite{Page:1974he}, the situation is different, the dominant thermal radiation leads to that the polarized directions of light waves are disorder so that the disk becomes a source of natural light without the total polarization. Thus, in the simulations of the polarized image of a black hole, only the hot disk model is considered. Moreover, as the linearly polarized light passes through the outer magnetized regions in plasma, it further undergoes the Faraday depolarization effects \cite{2000ApJ...538L.121A,Marrone:2005ky,Kuo:2014pqa}.

Along the path of each light ray from plasma to observer, the polarization components expressed by the Stokes parameters $(I, Q, U, V )$ \cite{1996A&AS..117..161H,2011A&A...527A.106S}
satisfy the polarized radiative transfer equations \cite{Krolik:2004ay,Shcherbakov:2008rg,Shcherbakov:2010kh,Dexter:2016cdk,Kulkarni:2011cy}
\begin{eqnarray}
\frac{\rm d\mathcal{I}}{\rm d\lambda}=\mathcal{J}-\mathcal{K}\mathcal{I},
\end{eqnarray}
where $\lambda$ is an affine parameter. The Stokes vector $\mathcal{I}=g^{3}(I, Q, U, V )$, the propagation matrix $\mathcal{K}$, and the emission vector $\mathcal{J}$ describe synchrotron emission and absorption coefficients in all Stokes
parameters, as well as Faraday rotation and conversion.
Thus, the propagations of the polarized light rays depend heavily on the plasma properties.

In the general relativistic magnetohydrodynamic (GRMHD) simulations, the plasma in the hot disk around the supermassive black hole can be simplified by a model, where the plasma is assumed to be collisionless with electrons and ions so that the electron temperature $T_e$ deviates from the
ion temperature $T_{i}$. The ratio
between the temperatures $T_{i}$ and $T_e$ can be expressed as \cite{Moscibrodzka:2015pda,Tsunetoe:2020pyz,Tsunetoe:2020nws}
\begin{eqnarray}
R=\frac{T_i}{T_e}=R_{\rm high}\frac{\beta^2}{1+\beta^2}+R_{\rm low}\frac{1}{1+\beta^2},
\end{eqnarray}
where $\beta$ is the ratio of gas pressure to magnetic pressure. $R_{\rm high}$ and $R_{\rm low}$  are numerical constants, which correspond to the ratio of ion to electron temperatures in the inner disk and in the jet region, respectively.

Through quantitatively evaluating a large library of images based
on GRMHD models and comparing  with the resolved EHT 2017 linear polarization map of
M87$^{*}$ \cite{2021ApJ...910L..12E,2021ApJ...910L..13E},  the viable GRMHD models revealed that the characteristic parameters for average intensity-weighted plasma  in the emission region are the electron number density
$n_e \sim10^{4-5} cm^{-3}$, the magnetic field strength $B\simeq7-30 G$, and the dimensionless
electron temperature $\theta_e \sim 8-60$.

Moreover, recent theoretical investigation shows that the polarization images of M87 jets are very sensitive to the black hole spin \cite{Tsunetoe:2020pyz}, which could provide a new possibility for measuring the spin parameter of a black hole. In the low-spin case,
there are much more symmetric ring shape patterns. This is
because the beaming and de-beaming effects are not so
large and the jet acceleration is not so significant
as the spin is small. In the high-spin case as $a=0.99M_{\rm BH}$, the polarized image
of the approaching jet disappeared in the low-spin case is clear \cite{Tsunetoe:2020pyz}. This is because the high black hole spin gives arise to that the
particle motion in the plasma can be accelerated up to the Lorentz factor of
$\Gamma_L\sim3$ and further yields that the approaching jet is more
bright than the counter one \cite{Tsunetoe:2020pyz}. Furthermore, there is the crescent-like image produced by the toroidal motion of gas blobs, which demonstrates that the jet acceleration process strongly depends on the black hole spin \cite{Nakamura:2018htq}.

One can also extract information about circular polarization through analyzing the Stokes quantity $V$ in the black hole images.
The circular polarization  can be amplified by the Faraday conversion in the well-ordered magnetic field.  This is different from the case of  the linear polarization where the polarization vectors are disordered by the strong Faraday rotation near the black
hole. Generally, in a model with hot disk, the circular polarization light images are faint and turbulent because the hot region occupied with chaotic magnetic fields is Faraday thick so that the Faraday conversion cannot be efficient. However, the study of circular polarization images
is helpful to understand  the polarized information in black hole images more completely. The combination of linear  and circular polarizations in future observations could provide a higher-precision detection on the magnetic structure, the temperature distribution and
the coupling between proton and electron near black holes.
It is shown that the circular polarization  images are
sensitive to the inclination angle \cite{Tsunetoe:2020nws}. Moreover, there is
a ``separatrix" in the  circular polarization images and across which the sign of the circular polarization is reversed. This can be attributed to the helical magnetic field structure in the disk \cite{Tsunetoe:2020nws}.
It implies that future full polarization EHT images are quite useful tracers of the magnetic
field structures near black holes.

The numerical simulations for the polarization image of the black hole are generally computationally expensive due to the broad parameter surveys and the complicated couplings among astrophysical and relativistic effects. Recently, a simple model of an equatorial ring of magnetized
fluid  has been developed to investigate the polarized images of synchrotron emission around the Schwarzschild
black hole \cite{EventHorizonTelescope:2021btj} and the Kerr black hole \cite{Gelles:2021kti}. Although only the emission from a single radius is considered,
this model can clearly reveal the dependence of the polarization signatures on the magnetic field configuration, the black
hole spin and the observer inclination. Moreover, with this model, the image of a finite thin disk can be produced by simply summing contributions from individual radii. The studies \cite{EventHorizonTelescope:2021btj,Gelles:2021kti}
also indicates that the ring model image is broadly consistent with the polarization
morphology of the EHT image. However, one must note that this  simple ring
model produces a high fractional polarization ($\geq 60\%$) even after blurring, which is much larger than that in the M87$^{*}$ image where the resolved fractional
polarization is about $\leq 20\%$ \cite{EventHorizonTelescope:2021btj}. This suggests that the significant depolarization
from the internal Faraday effects is essential when modeling
and interpreting the M87$^{*}$ image \cite{Moscibrodzka:2017gdx}. Nevertheless, the success
of the ring model in reproducing the structure of some
GRMHD images that have significant Faraday effects is encouraging
for the prospects of physical inference from this
simple model. Moreover, this simple model can be used to study the loops in the Stokes  $Q-U$ plane, which describes the continuous variability in the polarization around a black hole \cite{Eckart:2006fc,Trippe:2006jy,Zamaninasab:2009df,Fish:2009ak,Broderick:2005jj,Broderick:2005my}. It is beneficial to understand some time-varying features of emission from a localized orbiting hotspot near black hole in the real astronomical environment. Thus, this model has been recently applied
to study the polarized images of black holes in various spacetimes \cite{Qin:2021xvx,Zhu:2022amy,Delijski:2022jjj,Qin:2022kaf,Liu:2022ruc}.

In this simple ring model, the calculation of the polarization vector usually resorts to a so-called Walker-Penrose quantity \cite{Walker:1970un,Himwich:2020msm}. It is conserved along the null geodesic in the spacetimes where the dynamical system of photon motion is integrable and the equation of motion is full variable separable \cite{Walker:1970un}. The conserved Walker-Penrose quantity builds a direct connection between the polarization vectors of photon starting from the emitting source and reaching the observer. So in such spacetimes, the propagation of polarization vectors can be calculated by analytical methods, which greatly simplifies the calculation of polarization vectors along null geodesics.
However, in the spacetimes where the system of photon motion is nonintegrable, such as, in the Bonnor black dihole spacetime \cite{1966ZPhy..190..444B}, the Walker-Penrose quantity is no longer conserved along null geodesic.
Without the help of the Walker-Penrose constant, the calculation of the polarization vectors in this ring model  may still rely on the numerical methods. In the Bonnor black dihole spacetime,  there exist some fine fractal structures in the distribution of Stokes parameters $Q$ and $U$ in the polarized images \cite{Zhang:2022klr}. The signs of $Q$ and $U$ are opposite for two adjacent indirect images. It could be caused by that the photons forming two adjacent indirect images are emitted from the upper and
lower surfaces of accretion disk, respectively, resulting in a
large difference in the corresponding polarization vectors.

\section{Application prospects of black hole images}

The significance of studying black hole images lies in the following aspects. Firstly, such detections can
identify black holes and further verify and test the theories of gravity including general relativity,
and deepen our understandings on the nature of gravity. Secondly, analyzing information carried in black hole images
enables us to understand matter distribution and physical processes around the black holes, and
to give further insight into some fundamental problems in physics. In the following, we present some potential application prospects of black hole images.

\subsection{Probe the matter distribution around black holes}

To probe the matter distribution around black holes, one must simulate images of black hole models by considering different choices and select a model that could accurately represent the main features of the observed images. For the black hole M$87^{*}$, it is well known that it belongs to the class of low luminosity active galactic nuclei, and its spectral energy distribution presents features associated with emission from an optically thin and geometrically thick accretion disk ascribed to the synchrotron radiation with an observed brightness temperature in radio wavelengths in
the range of $10^9-10^{10}K$ \cite{2019ApJ...875L...1E}. Recently, the most salient features appearing in the EHT
Collaboration images of M$87^{*}$ were reproduced with impressive fidelity  and
the corresponding configuration model revealed that there may exist an asymmetric bar-like structure attached to a two-temperature thin disk in the equatorial plane of the black hole \cite{Boero:2021afh}. Moreover, the
asymmetry in brightness is a robust indicator of the orientation
of the spin axis. The simulations using different orientations of the black hole spin show that the spin direction opposite to the observed jet
is favored by the asymmetric shape of the observed crescent sector.

As mentioned in the previous part, the comparison between the polarization patterns of the M87$^*$ image and the viable GRMHD models reveals  the existence of magnetic field near the black hole. Actually, the magnetic field can generate some features of black hole images \cite{Wang:2021ara,Junior:2021dyw}. For a rotating black hole  immersed in a Melvin magnetic field \cite{Wang:2021ara},  the shadow becomes oblate for the weak magnetic field. However, in the case with the strong magnetic field, the multiple disconnected shadows emerge, including a middle oblate shadow and many striped shadows. Moreover, the novel feature in the Melvin-Kerr black hole shadow is the gray regions on both sides of the middle main shadow \cite{Wang:2021ara}, which are caused by the stable photon orbits around the stable light rings. In fact, the photons moving along the stable photon orbits are trapped and they can't enter the black hole. Strictly,
the gray regions don't belong to the black hole shadow, but if there are no light sources in the stable photon orbit regions, the observer also see dark shadows in the gray regions \cite{Wang:2021ara,Junior:2021dyw}. The chaotic lensing arising from the magnetic field gives rise to the self-similar fractal structures in the black hole shadows. The chaotic image also occurs for the case illuminated by an accretion disk in the Kerr-Melvin black hole spacetime with a strong enough magnetic field \cite{Hou:2022eev}. These new effects in shadows could provide a new way to probe the magnetic field near black holes.

The images of black holes indicate that the supermassive black holes in the centers of galaxies are actually surrounded by plasma. Besides as a light source to illuminate black holes, plasma is a dispersive medium where the index of refraction depends on
the spacetime point, the plasma frequency and  the photon frequency, so the plasma changes  the path of the light traveling through it and further affects the geometrical features of black hole shadows \cite{Perlick:2015vta,Atamurotov:2015nra,Abdujabbarov:2016efm,Crisnejo:2018uyn,
Huang:2018rfn,Yan:2019etp,Babar:2020txt,Matsuno:2020kju,Tsupko:2021yca,Li:2021btf,Atamurotov:2021cgh,Wang:2021irh,PhysRevD.104.084055,
Zhang:2022osx,Atamurotov:2021hoq,Perlick:2017fio,PhysRevD.92.104031,Bezdekova:2022gib}. The influence of plasma on the shadows
depends mainly on the ratio between the plasma frequency and
the photon frequency. If the plasma frequency is smaller than the photon frequency, the shadow is not very much different from the vacuum case. However, if the plasma frequency tends to the photon frequency, the significant changes in the photon regions will lead to a drastic modification of the properties of the shadow.
In the realistic case where the plasma frequency is much smaller than the photon
frequency, the plasma has a decreasing effect on the size of the shadows if the plasma density is higher
at the photon sphere than at the observer position. The above analyses are based on an assumption of plasma with radial power-law density. Recent study of angular Gaussian distributed plasma \cite{Zhang:2022osx}, where the plasma is non-spherically symmetric, shows that
the effect of plasma can be qualitatively explained by taking the plasma as a convex lens with the
refractive index being less than 1.
For the supermassive black holes at the centers of the Milky Way and the galaxy M87, which are the main targets of the current observations by the EHT, it is shown that the plasma effects start to become relevant at radio wavelengths of a few centimeters or more.
However, the present and planned instruments focus on the submillimeter range, where the scattering and
self-absorption have no significant effect on the emitted radiation around the
black holes and the plasma effects are very
small \cite{Perlick:2017fio,PhysRevD.92.104031}, so a realistic observation of the plasma influence on the shadows seems unfeasible at present.

\subsection{Constrain black hole parameters and test theories of gravity}

It is natural to expect to constrain black hole parameters by the using of shadows because the shape
and size of shadows depend on the black hole parameters
themselves.
In general, since black hole shadows have complex shapes in the observer's sky, the precise description of the shadow boundaries is crucial for measuring black hole
parameters.
\begin{figure}[htb]
\includegraphics[width=4cm ]{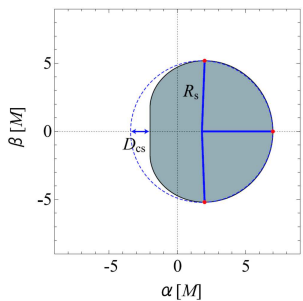}
\caption{The observables for the apparent shape of the Kerr black
hole $R_s$ and $\delta_s=D_{cs}/R_s$ \cite{2009PhRvD..80b4042H}.}
\label{shadowidentify}
\end{figure}
To fit astronomical observations, several observables were constructed by using special points
on the shadow boundaries in the celestial coordinates. For the Kerr black hole,
the two observables $R_s$ and $\delta_s=D_{cs}/R_s$ ( as shown in Fig.\ref{shadowidentify}) are introduced to measure the approximate size of the shadow and its deformation with respect to the reference circle \cite{2009PhRvD..80b4042H}, respectively. If the inclination angle is given, the values of the mass and spin of the black hole can be obtained by the precise enough measurements of $R_s$ and $\delta_s$. Recently, the length of the shadow boundary and
the local curvature radius are introduced to describe the shadow boundary \cite{Wei:2019pjf}. The black hole spin
and the observer inclination can be constrained  by simply measuring the maximum and minimum of the
curvature radius. Moreover, a topological covariant quantity is analyzed to measure and distinguish
different topological structures of the shadows \cite{Wei:2018xks,Cunha:2020azh}.
To further describe the general characterization of the shadow boundaries, a coordinate-independent formalism \cite{2015MNRAS.454.2423A} is proposed where the shadow curves $R_{\psi}(\psi)$ are expressed in terms of Legendre
polynomials  $R_{\psi}=\sum\limits_{l=0}^{\infty}c_lP_l(\cos\psi)$ with  the
expansion coefficients $c_l$. The dimensionless deformation parameters $\delta_{n}$ are defined to measure the relative
difference between the shadow at $\psi=0$ and at other angles $\psi=\pi/n $, $n=1,2,...k$, and $k$ is an arbitrarily positive integer.
These distortions are both accurate and robust so they can also be implemented to analyse the noisy data.

Above analyses are based on an assumption that the black hole shadows are
cast by a bundle of photons in parallel trajectories
that originating at infinity. For a realistic black hole surrounded by an accretion disk, the shadow is imprinted
on the image of the accretion flow. In principle, comparing a detailed model of the accretion disk around the black hole with
astronomical observations will yield a measurement of the size
and shape of the shadow. However, it is not feasible to predict the details of the brightness profile of the accretion flow image.
The first reason is the incompleteness of accretion disk models, and all theoretical models are simplified by introducing some assumptions so they are impossible to be completely consistent with the real disks.
The other reason is the observed
variability of the emission in the disk, since the inner accretion flow is highly turbulent and variable in the real astronomical environment.
Thus, it is necessary to build a procedure to analyze the observation data that focuses
on directly measuring  the properties of the shadow in a manner that is not seriously affected by our inability to
predict the brightness profile of the rest of the image \cite{Psaltis:2014mca}.
The gradient method \cite{1980RSPSB.207..187M,Canny1983FindingEA} is such kind of model-independent algorithms in image processing, which has already been applied successfully to interferometric images to quantify the properties of the turbulent structure of the interstellar magnetic field.
The basic concept in this algorithm is that the magnitude of the gradient of the
accretion flow image has local maxima at the locations of the steepest gradients, such as, in the case of the expected EHT images, which coincide with the edge of the back hole shadow \cite{Psaltis:2014mca}.  With the obtained gradient image where the rim of
the black hole shadow appears as the most discernible feature, a shadow pattern algorithm matching with the Hough/Radon
transform is employed to determine the shape and size of the shadow. This algorithm not only measures the properties of the black hole shadow, but also  assesses the statistical significance of the results.

The distinct features of black holes originating from deviation parameters in the alternative theory can help test the general relativity. It is shown that the shadow becomes prolate for the negative deviation parameter and becomes oblate for the positive one \cite{Broderick:2013rlq}.
The large deformation parameter in the Konoplya-Zhidenko rotating non-Kerr black hole yields the special cusp-shaped shadow for the equatorial observer \cite{Wang:2017hjl}. The large deviation arising from the quadrupole mass moment leads to chaotic shadow and the eyeball-like shadows with the self-similar fractal structures \cite{Wang:2018eui}. The similar features of shadows also appear in other non-Einstein theories of gravity including the quadratic degenerate higher-order scalar-tensor theories \cite{Long:2020wqj}.
Moreover, using the priori known estimates for the mass and distance
of M87$^{*}$  based on stellar dynamics \cite{2019ApJ...875L...1E,2019ApJ...875L...2E,2019ApJ...875L...3E,2019ApJ...875L...4E,2019ApJ...875L...5E,
2019ApJ...875L...6E,Gebhardt:2011yw,Walsh:2013uua,Kormendy:2013dxa}, the inferred size of the shadow from the horizon-scale images of the object M87$^{*}$ \cite{2019ApJ...875L...1E} is found
to be consistent with that predicted from general relativity for a Schwarzschild black hole within $17\%$ for a $68\%$ confidence interval.
However, this measurement still admits
other possibilities. The size of the black hole shadow
M87$^{*}$ can be used as a proxy to measure the deviations from Kerr metric satisfied weak-field tests \cite{EventHorizonTelescope:2020qrl}. For the parameterized Johannsen-Psaltis black hole, it has four lowest-order parameters and the shadow depends primarily on the parameter
$\alpha_{13}$ and only weakly on spin \cite{PhysRevD.87.124017}. The 2017 EHT measurement for M87$^{*}$ places a bound on the deviation parameter
$-3.6 < \alpha_{13} < 5.9$ \cite{EventHorizonTelescope:2020qrl}. For the modified gravity bumpy Kerr metric \cite{Gair:2011ym}, the size of the shadow depends primarily on the parameter
$\gamma_{1,2}$ and the requirement that the shadow size is consistent with the measurement of M87$^{*}$ within
$17\%$  gives a constraint on the deviation parameter $-5.0< \gamma_{1,2}< 4.9$ \cite{EventHorizonTelescope:2020qrl}. For the Konoplya-Rezzolla-Zhidenko
metric \cite{Konoplya:2016jvv}, the EHT measurements results in the constraint $-1.2< \alpha_{1}< 1.3$ \cite{EventHorizonTelescope:2020qrl}.
For these parametric deviation
metrics, the measurements  of the shadow size lead to significant constraints on the deviation parameters that
control the second post-Newtonian orders. This means that the EHT measurement of the size of a black hole leads to metric tests that are inaccessible in
the weak-field tests. In general, such parametric tests cannot be connected
directly to an underlying property of the alternative
theory. Recently, the EHT measurements have been applied to set
bounds on the physical parameters, such as,
the electric charge \cite{EventHorizonTelescope:2021dqv} and the MOG parameter in the Scalar-Tensor-Vector-Gravity Theory \cite{Kuang:2022ojj}. The quality of the measurements \cite{EventHorizonTelescope:2021dqv} is already sufficient to rule out that M87$^{*}$ is a highly charged dilaton
black hole, a Reissner-Nordstr\"{o}m naked singularity
or a Janis-Newman-Winicour naked singularity with large scalar charge.
Similarly, it also excludes considerable regions of the space of parameters for the doubly-charged dilaton and the Sen black holes.
Such tests are very instructive \cite{Amarilla:2010zq,Bambi:2008jg, Mizuno:2018lxz,Cunha:2019dwb} because they can shed light on which
underlying theories are promising candidates and which must
be discarded or modified.
The constraints and tests from shadows are complementary to those imposed by observations of
gravitational waves from stellar-mass sources.

\begin{figure}
\center{\includegraphics[width=8cm ]{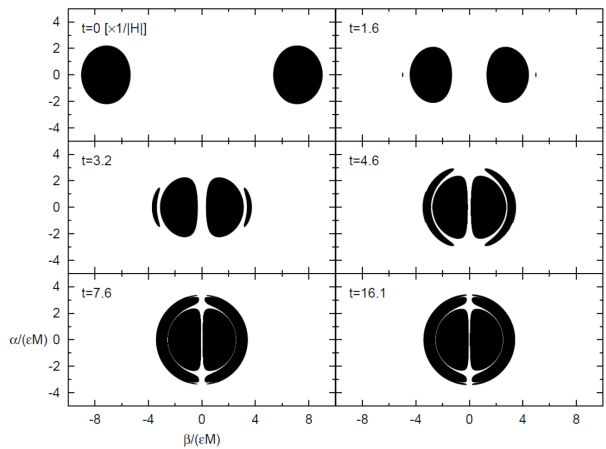}
\caption{The change of black hole shadows with time $t$ during the collision of two equal mass black holes\cite{Yumoto:2012kz}.}
\label{collid}}
\end{figure}

Black hole shadow may also provide a way to test binary black hole. Nowadays, the gravitational-wave events detected by the LIGO-Virgo-KAGRA Collaborations \cite{2016PhRvL.116f1102A,2016PhRvL.116x1103A,2017PhRvL.118v1101A,2017PhRvL.119n1101A,2017ApJ...851L..35A} confirm the existence of binary black hole system in the universe, and the systems of binary black hole are expected to be common astrophysical systems.
The shadows of the colliding between two black holes were simulated by adopting the Kastor-Traschen cosmological multiblack hole solution, which describes the collision of maximally charged black holes with a positive cosmological constant \cite{Nitta:2011in,Yumoto:2012kz}. Fig.\ref{collid} shows the change of the shadows with time $t$ during the collision of the two black holes with equal mass. At $t=0$, the two black holes are mutually away enough and their shadows are separated. However,  each shadow is a little bit elongated in the $\alpha$ direction because of the interaction between the two black holes. At $t=1.6$, the eyebrowlike shadows appear around the main shadows. The eyebrowlike shadows can be explained  by a fact that light rays bypass one black hole of binary system and enter the other one. With the further increase of time, the eyebrowlike structures grow and the main shadows approach each other. Although not discernible in the figure, in fact there appear
the fractal structures of the eyebrows, i.e., infinitely many thinner eyebrows at the outer region of these eyebrows as well
as at the inner region of the main shadows \cite{Yumoto:2012kz}.
As time elapses, the interval between two black hole shadows becomes indefinitely narrower, and it is expected that the black hole shadows eventually merge with each other \cite{Yumoto:2012kz}. However, due to the special properties of Kastor-Traschen metric, the recent investigation also implies that there is no observer who will see the merge of black hole shadows even if the black holes coalesce into one \cite{Okabayashi:2020apz}.
Another important solution of binary black hole with analytical metric form is Majumdar-Papapetrou solution, which describes the geometry of two extremally charged black holes in static equilibrium where gravitational attraction is in balance with electrostatic repulsion.
The similar eyebrowlike shadows are found in the Majumdar-Papapetrou binary black hole system  \cite{Shipley:2016omi}.

Actually, these eyebrowlike shadows with fractal structures also appear in other binary black hole systems, such as, in the double-Schwarzschild and double-Kerr black hole systems \cite{Cunha:2018cof} in which two black holes are separated by a conical singularity.
These common key features imprinted in the shadows of binary systems, such as disconnected
shadows with characteristic eyebrows,  open up a new analytic avenue for exploring four dimensional
black hole binaries \cite{Hertog:2019hfb}.

\subsection{Fundamental problems in physics}

\textit{ Dark matter}  The nature of dark matter is one of the most important open
fundamental questions of physics.  Dark matter is assumed to be an invisible matter, which constitutes the dominant form of matter in the universe and has feeble couplings with the common visible matter at most.
Despite extensive observational data  supporting its presence on a large scale, dark matter has not been  directly detected  by any scientific instrument. Dark matter should influence black hole shadow due to its gravitational effects.
A simple spherical model consisting of a Schwarzschild black hole with mass $M$ and a homocentric spherical shell
of dark matter halo with mass $\Delta M$ is applied to tentatively study the effects of dark matter on the black hole shadow \cite{Konoplya:2019sns}. It is found that the mass of dark matter and its distance over mass distribution lead to larger radius of shadows. However, it must be pointed out that
in this simple model the dark matter is unlikely to manifest itself in the shadows of galactic black holes, unless its concentration near black holes is abnormally high \cite{Konoplya:2019sns}.

The effect of dark matter halo on black hole shadows has been studied in the spacetimes  of a spherically symmetric black hole and of a rotating black hole \cite{Hou:2018bar,Xu:2018wow,Jusufi:2020cpn,Jusufi:2019nrn,Cunha:2019ikd}. It is shown that the structures of the black hole shadows in the cold dark matter (CDM) and scalar field dark matter (SFDM) halos
are very similar to the cases of the Schwarzschild and Kerr black holes, respectively. Both dark matter models
influence the shadows in a similar way and the sizes of the shadows increase
with the dark matter parameter $k\equiv\rho_c R^3$, where the characteristic density $\rho_c$ and the radius $R$ are related to the distribution of dark matter halo in two models. In general, the influence
of the dark matter on the black hole shadows is minor and only becomes significant when $k$ increases to
order of magnitude of $10^7$ for both CDM and SFDM models \cite{Hou:2018bar}. The calculation
of the angular radii of the shadows shows that the dark matter halo could influence the
shadow of Sgr A$^{*}$ at a level of order of magnitude of $10^{-3} \mu as$ and $10^{-5} \mu as$, for CDM and
SFDM, respectively. However, it is out of the reach
of the current astronomical instruments \cite{Hou:2018bar}. The current EHT resolution is  $\sim60 \mu as$ at 230 GHz and will achieve $15 \mu as$ by observing at a higher
frequency of 345 GHz and adding more very long baseline interferometry (VLBI) telescopes. The space-based VLBI RadioAstron \cite{Kardashev:2013cla}
will be able to obtain a resolution of  $1-10 \mu as$. This is still at least three orders of magnitude lower than the resolution required by the CDM model. The black hole shadow has been studied for a rotating black hole
solution surrounded by superfluid dark matter and baryonic matter.  Using the current values for the parameters of the superfluid
dark matter and baryonic density profiles for the Sgr A$^{*}$ black hole,  it is shown that the effects of the
superfluid dark matter and baryonic matter on the sizes of shadows are almost negligible compared
to the Kerr vacuum black hole \cite{Jusufi:2020cpn}. Moreover, comparing with the dark matter, the shadow
size increases considerably with the baryonic mass. This can be understood by the fact that the baryonic matter is mostly located in the galactic center. Similarly, the baryonic matter in this model yields an increase of the angular diameter of the shadow of the magnitude
$10^{-5} \mu as$ for the Sgr A$^{*}$ black hole  \cite{Jusufi:2020cpn}.

The axion is a hypothetical particle beyond the standard
model, which is initially proposed  to solve the strong CP (charge-conjugation and parity) problem \cite{PhysRevLett.38.1440,PhysRevLett.40.223,PhysRevLett.40.279,DiLuzio:2020wdo}.
Nowadays, axionlike particles  are also introduced in fundamental theories and served as an
excellent dark matter candidate so there are many search
experiments designed to prob axions \cite{Marsh:2015xka,Sikivie:2020zpn,Raffelt:2006cw,Raffelt:1990yz,CAST:2017uph,Payez:2014xsa}.  Axion cloud around a rotating black hole may be formed through the superradiance
mechanism if the Compton wavelength of axion particle is at the
same order of the black hole size \cite{Strafuss:2004qc,Brito:2015oca}. Due to the existence of the axion cloud, the axion-electromagnetic-field coupling  gives rise to that the position angles of linearly polarized photons emitted near the horizon oscillate periodically \cite{PhysRevD.41.1231,Plascencia:2017kca,Fujita:2018zaj,Fedderke:2019ajk}. Along this line,
a novel strategy of detecting axion clouds around supermassive black holes is recently proposed by using the high spatial resolution and polarimetric measurements of the EHT \cite{Chen:2019fsq}.
 \begin{figure}
\includegraphics[width=8cm ]{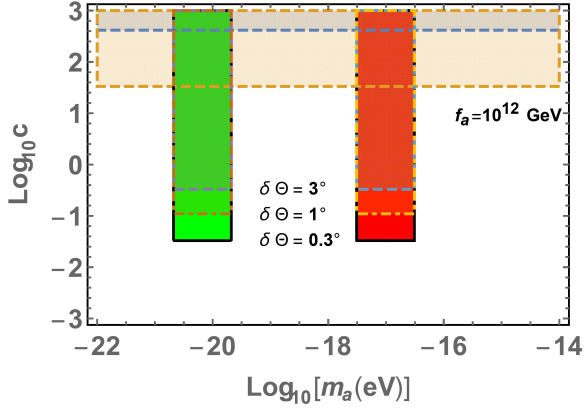}
\caption{The expected axion parameter space probed by polarimetric
observations of M87$^{*}$ (green) and Sgr A$^{*}$ (red) for different position
angle precisions \cite{Chen:2019fsq}. The bounds from  CAST \cite{CAST:2017uph} (gray) and Supernova 1987A (pale yellow) are shown to make a comparison.}
\label{axion}
\end{figure}
Fig.\ref{axion} presents the axion parameter space which is
potentially probed by M87$^{*}$ and Sgr A$^{*}$
for different position angle precisions \cite{Chen:2019fsq}. This
method is complementary to the constraints from the black
hole spin measurements through  gravitational wave detections \cite{Arvanitaki:2014wva}. Since the position angle oscillation
induced by the axion background does not depend on
photon frequency, it is expected that polarimetric measurements at different frequencies in the future can be used to distinguish astrophysical background and to improve the sensitivity of tests of the axion superradiance scenario. Moreover, the possibility of probing ultralight axions by the circular
polarization light is also studied in \cite{Shakeri:2022usk}.

\textit{Extra dimension} The possible existence of extra dimensions is one of the
most remarkable predictions of the string theory. The extra spatial dimension could play an important role in fundamental theories within the
context of the unification of the physical forces and also in black hole physics. For the high-dimensional
black holes, it is shown that the extra dimension influences the shape and size
of the shadows \cite{Hertog:2019hfb,Singh:2017vfr,Papnoi:2014aaa,Amir:2017slq}. Using the size
and deviation from circularity of the shadow of the black hole M87$^{*}$ observed by the EHT collaboration,  the  curvature radius of AdS$_5$ in the Randall-Sundrum brane-world scenario is bounded by an upper limit $l\lesssim170$AU \cite{Vagnozzi:2019apd}. This upper limit is far from being competitive
with current  $\mathcal{O}$ (mm) scale constraints from precision tests of gravity, but greatly improves
the limit $l\lesssim0.535 $ Mpc obtained from GW170817 \cite{Visinelli:2017bny}. More importantly, it is
an independent limit from imaging the dark shadow of M87$^{*}$.
Using a rotating black hole solution with a cosmological in the vacuum
brane,  the black hole shadow together with the observed data of M87$^{*}$ also provides a
upper bound  for the normalized tidal charge $q<0.004$ \cite{Neves:2020doc}, which is the second best result for the tidal charge to date and is a little higher than the best one $q<0.003$ from a solar system test \cite{Boehmer:2008zh}. Moreover, the negative values of the tidal charge are reported to be favored with the M87$^{*}$ and Sgr A$^{*}$ data in the brane contexts by the using of Reissner-Nordstr\"{o}m-type geometry \cite{PhysRevD.82.064009,Zakharov:2018awx,Horvath:2012ru} and a rotating black hole without a cosmological constant \cite{Banerjee:2019nnj}.

For the case of the compactified extra dimension, the shadow of a rotating uniform black string has been studied where the extra spatial dimension is treated as a compacted circle with the circumference $l$ \cite{Tang:2022hsu}. The momentum of photon arising from the fifth dimension enlarges the photon regions and the shadow of the rotating 5D black string while it has slight
impact on the distortion. The angular diameter
in the EHT observations of M87$^{*}$ leads to the constraint on the length of the compact extra
dimension 2.03125 mm $ \lesssim l\lesssim$ 2.6 mm \cite{Tang:2022hsu}. Similarly, from the observations of Sgr A$^{*}$, the constraints
2.28070 mm $ \lesssim l\lesssim$ 2.6 mm and 2.13115 mm  $ \lesssim l\lesssim$ 2.6 mm
can be given by the upper bounds of the emission ring and the
angular shadow diameter respectively \cite{Tang:2022hsu}. In particular, within these bounds,
the rotating 5D black string spacetime is free from the Gregory-Laflamme instability \cite{Tang:2022hsu}.

Effects of the specific angular
momentum $\xi_{\psi}$ of photon from the fifth dimension on black hole shadow have also been studied for a rotating squashed Kaluza-Klein black hole \cite{Long:2019nox}, which is a kind of interesting
Kaluza-Klein type metrics with the special topology and asymptotical structure \cite{Wang:2006nw}. It has squashed $S^3$ horizons so the black hole
has a structure similar to a five-dimensional black hole in the vicinity of horizon, but behaves as the four-dimensional
black holes with a constant twisted $S^1$ fiber in the far region. For this special black hole,
the radius $R_s$ of the black hole image in the observer's sky  has different values for the photons
with different angular momentum $\xi_{\psi}$. The real radius of the black shadow is equal to
the minimum value of $R_{s_{\rm min}}$. Especially, as
the black hole parameters lie in a certain special range, it is found that there is no shadow for
a black hole since the minimum value $R_{s_{\rm min}}=0$ in these special cases \cite{Long:2019nox}, which is novel since
it does not appear in the usual black hole spacetimes. It must be pointed out that the emergence of black hole without shadow
does not mean that light rays can penetrate through the black hole. Actually, it is just because the photons near the black hole with certain range of $\xi_{\psi}$ change their propagation
directions and then become far away from the black hole.  The phenomenon of black hole
without black shadow will vanish if there exists the further constraint on the specific
angular momentum $\xi_{\psi}$  of photon from the fifth dimension. In the case where black hole
shadow exists, the radius of the black hole
shadow increases monotonically with the increase of extra dimension parameter in the non-rotating case.
With the increasing of rotation parameter, the radius of the black hole shadow gradually becomes a monotonously decreasing function of the extra dimension parameter. With the latest observation
data, the angular radii of the shadows for the supermassive black hole Sgr A$^{*}$
at the centre of the Milky Way Galaxy and the supermassive black hole in M87 are estimated \cite{Long:2019nox},
which implies that there is a room for the theoretical model of such a rotating squashed Kaluza-Klein black hole.

\textit{Coupling between the photon and background field} Analogous to the motion of charged particles in an electromagnetic field, the propagation of light rays in a spacetime is also influenced by the coupling between the photon and background field, which could leave observable effects on the
black hole shadow. In the standard Einstein-Maxwell theory, there
is only a quadratic term of Maxwell tensor directly related
to electromagnetic field, which can be seen as
an interaction between Maxwell field and  metric tensor.
Actually, the interactions between electromagnetic field and
curvature tensor could appear naturally in quantum
electrodynamics with the photon effective action originating
from one-loop vacuum polarization \cite{Drummond1980QEDVP}. Although
these curvature tensor corrections appear firstly as an effective
description of quantum effects, the extended theoretical models without
the small coupling constant limit have been investigated for
some physical motivations \cite{Hehl1999HowDT,Balakin:2007am,Clarke:2000bz,Bamba:2008ja}.

The coupling between the photon and Weyl tensor leads to birefringence phenomenon so that the paths of light ray propagations are different for the coupled photons with different polarizations. Thus, it is natural to give rise to double shadows for a single black hole because the natural lights near the black hole can be separated into two kinds of linearly
polarized light beams with mutually perpendicular polarizations \cite{2016EPJC...76..594H}.
With the increase of the coupling strength, the umbra of the black hole decreases and the penumbra increases.
In the case of an equatorial thin accretion
disk around the Schwarzschild black hole, the black hole image and its polarization distribution are also affected by the coupling strength \cite{Zhang:2021hit}.
The observed polarized intensity in the bright region is
stronger than that in the darker region. It is also noted that
the effect of the coupling on the observed polarized vector is weak in
general and the stronger effect  appears in the bright
region close to the black hole in the image plane. Moreover, for the different coupling strengths, the observed polarized patterns have a counterclockwise vortex-like distribution with a rotational symmetry as the observed inclination angle $\theta_0=0^{\circ}$. The rotational symmetry in polarized patterns gradually vanishes with the increase of the inclination angle.
Quantum electrodynamic  effects  from  the  Euler-Heisenberg effective Lagrangian on the shadow have been studied in the black hole background \cite{Hu:2020usx}. Similarly, in this case, the birefringence effect also yields that observer sees different shadow sizes of a single black hole for different polarization lights.

The coupling between a photon and a generic vector field is also introduced to study black hole shadow \cite{Yan:2019hxx}. The generic vector field is
assumed to obey the symmetries possessed by the black hole and the boundary condition that the vector field vanishes
at infinity. It is found that the black hole shadow in edge-on view also has  different appearances
for different frequencies of the observed light. This is because the coupling form
alters the way that the system depends on the initial conditions. These new phenomena about the black hole shadow originating from
the coupling between the photon and background vector field are not simply caused  by modifications of the metric, which could help  give insight into new physics \cite{Yan:2019hxx}. In particular, such a kind of coupling can affect the motion of photons and phenomenologically depict a violation of equivalence principle \cite{Yan:2019hxx}. Thus, it
is proposed as a mechanism  to test the equivalence principle by analyzing black hole
shadows. Although  the current observation conditions might not allow us to directly detect these novel phenomena, it is expected that the future project of the next generation EHT  with other future multi-band observations \cite{Blackburn:2019bly} as well as the related data-processing techniques could allow for tests of
these new physics imprinted in the black hole shadows. Moreover, the shadow images of M87$^{*}$ and Sgr A$^{*}$ are recently used to constrain the  parameters in the generalized uncertainty principle (GUP) \cite{Neves:2019lio} and the Lorentz symmetry
violation \cite{Khodadi:2022pqh}, respectively. Although these best upper limits are weaker than those obtained in most other physical
frameworks, they are valuable for further understanding black hole images and fundamental problems in physics \cite{Vagnozzi:2022moj,Ozel:2021ayr,Psaltis:2018xkc}.

\section{Summary}

The near-horizon images of the shadows of the supermassive compact objects M87$^{*}$ and Sgr A$^{*}$ delivered by the EHT have opened an amazing window for the strong-field test of gravity theories as well as fundamental physics. These images are composed of black hole shadow and the image of accretion disk around the central black hole. Black hole shadow is essentially formed by the light rays entering
 the black hole's event horizon,  in spite that its shape and size also depend on the position of observer and the types of light sources. The fundamental photon orbits and the invariant phase space structures determine the intrinsic features of the black hole shadow.
However, the visualization of the shadow must resort to the emission in the accretion disk around the black hole in the real astronomical environment.
This means that the visible images of the black hole also depend on the properties of the accretion disk and the physical processes in the disk, which
yields that the black hole images could have a highly model-dependent appearance \cite{Psaltis:2014mca}. For example, some models
show a partially obscured shadow and others present an apparently
exaggerated shadow. Especially, if the disk is optically thick, there may be no visible shadow at all, which means that the geometrical
thickness is a key ingredient for observing the shadow.
On the other hand, the information on luminance and polarization stored in the image of accretion disk can be helpful to understand the matter distribution and structures in the strong field region near the black hole.

Although black hole shadow and image carry the characteristic information of a black hole, it must be pointed out that the black hole shadows and images in some spacetimes may be not sensitive enough to certain parameters  so that the effects of these parameters on the black hole images can not be discriminated in terms of the resolution of the current observation devices.  With the increasing accuracy and resolution of the future astronomical observations and the technological development, as well as the more theoretical investigations,
it is expected that these mint markings of black holes can be more clearly detected in the next generation EHT, the BlackHoleCam and the space-based experiments.
The future detections of the fractural fine structures in
black hole shadows arising from the chaotic lensing and the competitive constraints on fundamental physics principles from black hole shadows will help  better test theories of gravity and to deeply understand the fundamental problems in modern physics. In a word, the study of black hole images is still in its infancy, and the detection of images for M87$^{*}$ and Sgr A$^{*}$ black holes is only a starting point.

\section{\bf Acknowledgments}

We would like to thank Profs. Carlos Herdeiro and Jieci Wang for their useful comments and suggestions. This work was supported by the National Natural Science Foundation of China under Grant Nos. 12035005, 12275078 and
11875026.

\bibliography{black hole shadow}

\end{document}